\documentclass{article}

\usepackage{arxiv}

\usepackage[utf8]{inputenc} % allow utf-8 input
\usepackage[T1]{fontenc}    % use 8-bit T1 fonts
\usepackage{hyperref}       % hyperlinks
\usepackage{url}            % simple URL typesetting
\usepackage{booktabs}       % professional-quality tables
\usepackage{amsfonts}       % blackboard math symbols
\usepackage{nicefrac}       % compact symbols for 1/2, etc.
\usepackage{microtype}      % microtypography
\usepackage{graphicx}
\usepackage[numbers]{natbib}
\usepackage{doi}

\usepackage{caption}
\usepackage[dvipsnames]{xcolor}
\usepackage{colortbl}
\usepackage{xspace}
\usepackage{multirow}
\usepackage{csquotes}
\usepackage{algorithm}
\usepackage{algpseudocode}
\usepackage{longtable}
\definecolor{LightGray}{gray}{0.9}

\usepackage{mdframed}
\newcounter{finding}
\newcommand{\finding}[1]{\refstepcounter{finding}
 \begin{mdframed}[linecolor=gray,roundcorner=12pt,backgroundcolor=gray!15,linewidth=3pt,innerleftmargin=2pt, skipabove=10pt, skipbelow=10pt, leftmargin=0cm,rightmargin=0cm,topline=false,bottomline=false,rightline = false]
  \textbf{Finding \arabic{finding}} \textit{#1}
 \end{mdframed}
}
\newcounter{resarchq}
\newcommand{\resarchq}[1]{\refstepcounter{resarchq}
 \begin{mdframed}[linecolor=Cerulean,roundcorner=12pt,backgroundcolor=SkyBlue!15,linewidth=3pt,innerleftmargin=2pt, skipabove=10pt, skipbelow=10pt, leftmargin=0cm,rightmargin=0cm,topline=false,bottomline=false,rightline = false]
  \textbf{RQ\arabic{resarchq}} #1
 \end{mdframed}
}

\usepackage[accsupp]{axessibility}  % Improves PDF readability for those with disabilities.
\usepackage{cleveref}

\usepackage{todonotes, soul}

\newcommand{\freyr}{\raisebox{-1.5pt}{\includegraphics[height=1.05em]{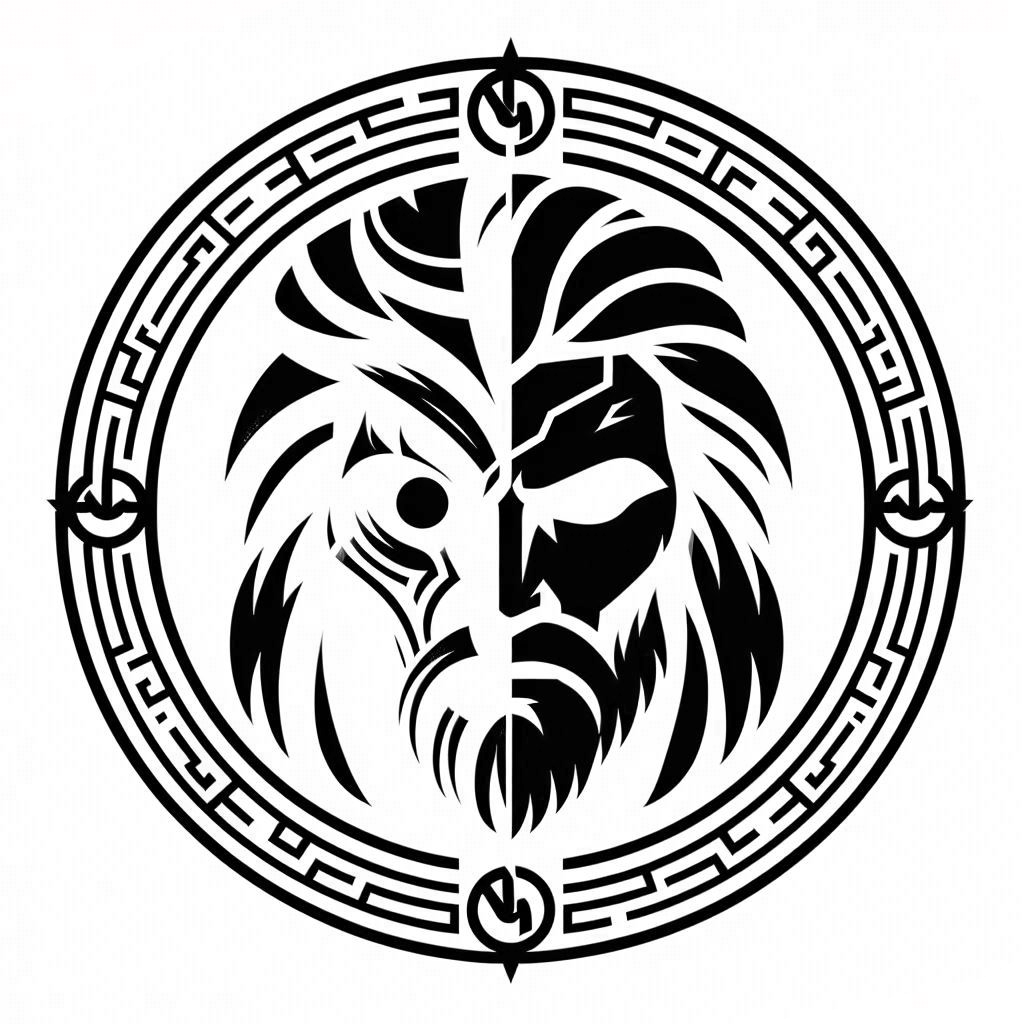}}\xspace}
\newcommand{\github}{\raisebox{-1.5pt}{\includegraphics[height=1.05em]{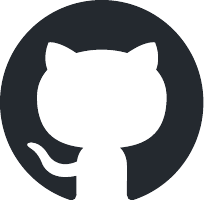}}\xspace}

\title{\freyr\textbf{Freyr}: A \textbf{F}ramework for \textbf{R}ecognizing and \textbf{E}xecuting \textbf{Y}our \textbf{R}equests}

\date{}

\author{
    \href{https://orcid.org/0000-0001-7578-6173}{\includegraphics[scale=0.06]{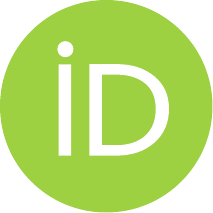}\hspace{1mm}Roberto Gallotta}\\
	Institute of Digital Games\\
	University of Malta\\
	L-Imsida, Malta\\
	\texttt{\href{mailto:roberto.gallotta@um.edu.mt}{roberto.gallotta@um.edu.mt}}\\
	\And
    \href{https://orcid.org/0000-0001-5554-1961}{\includegraphics[scale=0.06]{images/orcid.pdf}\hspace{1mm}Antonios Liapis}\\
    Institute of Digital Games\\
    University of Malta\\
    L-Imsida, Malta\\
    \texttt{\href{mailto:antonios.liapis@um.edu.mt}{antonios.liapis@um.edu.mt}}\\
	\And
    \href{https://orcid.org/0000-0001-7793-1450}{\includegraphics[scale=0.06]{images/orcid.pdf}\hspace{1mm}Georgios N. Yannakakis}\\
    Institute of Digital Games\\
    University of Malta\\
    L-Imsida, Malta\\
    \texttt{\href{mailto:georgios.yannakakis@um.edu.mt}{georgios.yannakakis@um.edu.mt}}\\
}

\renewcommand{\shorttitle}{\textsc{Freyr}}

\hypersetup{
pdftitle={\textsc{Freyr}: A Framework for Recognizing and Exectuing Your Requests},
pdfsubject={},
pdfauthor={Roberto Gallotta, Antonios Liapis, Georgios N. Yannakakis},
pdfkeywords={Large Language Model, Tool Usage, Natural Language Understanding},
}

\begin{document}
\maketitle

\begin{abstract}
Large language models excel as conversational agents, but their capabilities can be further extended through tool usage, \textit{i.e.}: executable code, to enhance response accuracy or address specialized domains. Current approaches to enable tool usage often rely on model-specific prompting or fine-tuning a model for function-calling instructions. Both approaches have notable limitations, including reduced adaptability to unseen tools and high resource requirements. This paper introduces \textsc{Freyr}, a streamlined framework that modularizes the tool usage process into separate steps. Through this decomposition, we show that \textsc{Freyr} achieves superior performance compared to conventional tool usage methods. We evaluate \textsc{Freyr} on a set of real-world test cases specific for video game design and compare it against traditional tool usage as provided by the Ollama API.

\begin{center}
    \begin{tabular}{rl}
     \github & \href{https://github.com/gallorob/freyr}{\path{gallorob/freyr}} \\
    \end{tabular}
\end{center}
 
\end{abstract}


\begin{figure*}[h!]
    \centering
    \includegraphics[width=0.8\textwidth]{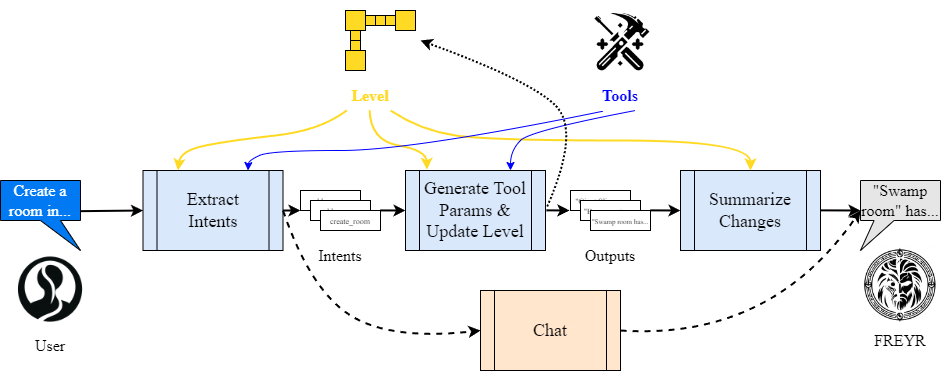}
    \captionsetup{labelformat=empty}
    \caption{An overview of our \textsc{Freyr} framework on editing a video game level.}
    \label{fig:freyr_diagram}
\end{figure*}
\addtocounter{figure}{-1}

\section{Introduction}
Large language models (LLMs) have become an essential part of numerous applications, ranging from domain-specific chatbots \cite{dan2023educhat,he2024physician,yang2024quokka} to code generation \cite{acher2023programming,gao2022pal,mallick2024chatvis,ross2023programmer} and content creation \cite{baek2024chatpcg,taveekitworachai2024chatgpt4pcg,xia2024llmga}. While their ability to generate text is remarkable, LLMs can be extended beyond these interactions through tool usage. Tools allow LLMs to execute functions, run code, or interact with external systems, enabling them to not only produce responses but also perform computations, generate content, or interface with specialized domains. This capability makes LLMs that support tool usage significantly more versatile, allowing them to reason and act directly \cite{ahn2022saycan,huang2022language,nakano2022webgpt}.

Not all LLMs however support tool usage. Current approaches to enabling tool usage often fall into two main categories. The first involves instruction prompting, where a description of how to format a response is given to the LLM. The response is then parsed to execute different predefined tools. This method is model-specific and lacks versatility across different tools and scenarios, but can handle a dynamic set of tools presented to the model at runtime. The second category involves fine-tuning the model to condition the response format to interact with a predefined set of tools. This method improves performance but requires significant resources to fine-tune the model and reduces the model's generalizability. Moreover, both approaches face challenges when a large number of tools is given to the model at once. As tools are described to the LLM as text, they must fit in the model's input context size. Open-source models often are unable to fully utilize their full context length \cite{li2023how}, and even large proprietary model's performance starts degrading after 10 to 20 tools \cite{openaitoolslownum}.

To address these limitations, we propose a \enquote{Framework for Recognizing and Executing Your Requests} (\textsc{Freyr}). \textsc{Freyr} falls into the first category of approaches for tool usage, however it modularizes the pipeline for generating responses via tools into discrete, explicit steps. By decomposing the pipeline into its core phases, \textsc{Freyr} enhances the efficiency and reliability of tool-enabled LLMs. This modular approach allows for greater flexibility in adapting to new tools and reduces reliance on model-specific or resource-intensive methods. Furthermore, \textsc{Freyr} leverages consumer-level open-source LLMs, ensuring compatibility with local environments and avoiding third-party dependencies. This makes \textsc{Freyr} an accessible solution for a broader range of users.

We evaluate \textsc{Freyr} on the \textit{LLMaker} test set, a content creation domain where the LLM is tasked to iteratively refine video game assets following a designer's requests. We compare \textsc{Freyr} performance against readily-available tool usage provided by the popular Ollama API \cite{ollama}. 

We release the full implementation of \textsc{Freyr} on GitHub at \url{https://github.com/gallorob/freyr}. This open-source release includes the full codebase to reproduce and evaluate our results, and allows extensions of the base framework.

\section{Related Work}
In this work we introduce a novel approach to allow LLMs to use tools (\Cref{subsec:tool_usage}) that leverages intent recognition (\Cref{subsec:intent_rec}). We test our approach on the test cases first introduced in \textit{LLMaker} (\Cref{subsec:llmaker}).

\subsection{Tool Usage}\label{subsec:tool_usage}
The ability of LLMs to use external tools has recently garnered much attention. Tool usage allows LLMs to integrate external knowledge into their responses \cite{chen2024octopus,patil2023gorilla}, or even perform tasks that go beyond pure text generation \cite{gallotta2024consistent}. Enabling LLMs to use tools initially required collecting a large data set of human annotated tool interactions, which was resource intensive and time consuming. This was the approach used in \textit{Toolformer} \cite{schick2023toolformer}. More recent work instead focused on generating such datasets using other LLMs, such as in \textit{GPT-4Tools} \cite{yang2023gpt4tools}.

One prominent application of tool usage for LLMs is generating and executing API calls, which are functionalities provided to clients by a server or application. This can be problematic for LLMs as there can be thousands of possible functions to choose from \cite{chen2024octopus,tang2023toolalpaca}, requiring smart ways to only provide a handful of tools to the LLM, so as not to exceed the LLM's context length. \textit{Gorilla} \cite{patil2023gorilla}, for example, leveraged a retrieval-augmented generation (RAG) approach to select relevant tools, whereas \textit{ToolPlanner} \cite{liu2024toolplanner} clustered tools by similarity.

These existing frameworks however are limited because they rely on static fine-tuned models or heuristic-driven prompt templates. The popular Ollama API platform instead enables tool usage by ad-hoc prompting each of the officially supported LLMs\footnote{All officially supported models are listed at \url{https://ollama.com/search?c=tools}.}. \textit{ReAct} \cite{yao2022react} instead prompted LLMs to generate interleaved verbal reasoning traces between tool calls, which improved tool usage accuracy. However these approaches still struggle to correctly generate the correct response when presented with a large number of diverse tools. This is because they require the full description of the tools. The framework we propose is instead more scalable, as the choice of tools to execute requires a shorter description to include in the prompt. Additionally, we target existing off-the-shelf LLMs to correctly execute tool calls without the need for a RAG system, further reducing the need for resources both on train and deployment of the system.

\subsection{Intent Recognition}\label{subsec:intent_rec}
Humans are very good at expressing the same concept in a myriad different ways \cite{pinker2008though}, which makes it very difficult to build computing systems that are able to correctly identify and process requests. The core idea behind intent recognition is learning a mapping between human expressions and grounded, distinct intents. This mapping can be learned by deep learning models \cite{casanueva2020efficient}, which however can suffer from lack of adaptability and robustness to unseen tools. More recent advances improved intent recognition using relationship meta-features \cite{siddique2021generalized}, and leveraging the popular transformer architecture \cite{lamanov2022template, wang2024intelligent}. These methods excel in structured environments but often require substantial pre-training or, as black-box models, lack explainability.

With the surge in popularity of LLMs, recent work has demonstrated their efficacy in the task of intent recognition. LLM agents can predict intents from a pre-defined set \cite{kim2024autointent}, or leveraging external knowledge bases \cite{taju2023chatbot}. Fine-tuning LLMs for intent detection specifically has also been explored with promising results \cite{lina2024finetuning}. Acting upon detected intents also improves LLMs explainability, as highlighted by Wang \textit{et al.} \cite{wang2024llmcheckup}.

In this work, we rely on intent detection to interpret user requests, leveraging the LLM's commonsense implicit knowledge and providing a set of possible intents in the prompt. This type of in-context learning requires less resources than fine-tuning a model, and can overcome more rigid classification system and heuristic-driven intent mappings.

\subsection{LLMaker}\label{subsec:llmaker}
\textit{LLMaker} \cite{gallotta2024llmaker} is a mixed-initiative content creation tool that utilizes LLMs to assist human users in designing video game levels for the reverse dungeon crawler game \textit{Dungeon Despair}. Inspired by \textit{Darkest Dungeon} (Red Hook Studios, 2016), the game features heroes battling enemies within rooms and corridors of a dungeon.

In their introductory paper, Gallotta \textit{et al.} \cite{gallotta2024consistent} compared different prompting methods to edit the game level, evaluated on GPT-3.5-Turbo 1106. Using a set of test cases that mimicked a designer interacting with the tool, Gallotta \textit{et al.} determined that function calling was the best performing approach. At its core, \textit{LLMaker} relies on a predefined set of functions that are called through natural language instructions provided by the human user. During design, the LLM uses context and domain constraints to generate function names and necessary arguments for each call. These functions impact the game's level by adding, editing, or removing rooms, corridors, enemies, traps, or treasures. Once a call is completed, the LLM provides a brief summary of the changes to the user. If a function fails to execute, a functional error \cite{buonanno2022functional} is instead returned to the LLM, which can then decide whether to retry calling the function with different parameters or inform the user of the issue.

\section{A \texttt{F}ramework for \texttt{R}ecognizing and \texttt{E}xecuting \texttt{Y}our \texttt{R}equests}\label{sec:freyr}
As introduced in \Cref{subsec:tool_usage}, LLMs can use tools by generating a response that contains the tool name and the required parameters. Responses generated via tools require at least two LLMs inference passes: the first to generate the tool call parameters, and the second to report to the user the result of the tool call. However, this implies that all tools properties are available to the LLM at once, significantly increasing the number of tokens present in the prompt. While this approach is ideal for use cases where a small number of tools is available to the LLM, it does not scale well with multiple tools, requiring larger and larger context windows.

In this work we instead suggest that the tool calling pipeline can be split into separate steps, requiring at best only one additional inference step. Let us imagine that an LLM receives a request from the user and has only one tool at its disposal: the first thing it must do is decide whether a tool should be used. This first step has been explored explicitly for RAG applications, letting the model decide whether it should retrieve information from a knowledge base \cite{asai2024selfrag}. If not, the LLM generates a response in natural language. If instead a tool should be used, the LLM must generate the parameters for the tool. It then receives the tool execution result, and provides a summary response in natural language to the user. \textsc{Freyr} integrates this pipeline explicitly with four separate modules: one to determine whether a tool should be used ($LLM_{intent}$), one to generate the tool parameters ($LLM_{parameters}$), one to summarize the tool outputs ($LLM_{summary}$), and one to process conversations ($LLM_{chat}$). An overview of the entire \textsc{Freyr} pipeline is presented in \Cref{alg:freyr}. We note that we do not enforce a structured generation of the outputs, instead specifying a more free-form response format (comma-separated or bulleted list) that we parse afterwards\footnote{All prompts used in this work are available at \url{https://github.com/gallorob/freyr/resources}.}. This \enquote{divide and conquer} approach allows us to select up to four different LLMs for each sub-task, giving great flexibility to the overall system and allowing for specific optimizations to take place. For example, $LLM_{chat}$ can be a very small LLM that is very good for conversation, while employing a larger LLM for the $LLM_{parameters}$ role that is more suitable for generating the parameters required by the tool. This is in direct contrast with current approaches, where the same LLM is used for the entire tool calling pipeline. 

\begin{algorithm}
\caption{\textsc{Freyr} algorithm overview.}
\label{alg:freyr}
\begin{algorithmic}[1]
    \Require{$LLM_{intent}$, $LLM_{parameters}$, $LLM_{summary}$, $LLM_{chat}$, $Tools$, $Conversation$, $Message$, $Level$, $max\_retries$}
    \Ensure{A $response$ generated for the user.}
    \State $intents \leftarrow LLM_{intent}(Conversation, Message, Tools, Level)$
    
    \If{$intents \equiv \{\mathrm{``conversation"}\}$}
        \State $response \leftarrow LLM_{chat}(Conversation, Message, Level)$
    \Else
        \State $outputs \leftarrow \emptyset$
        \For{\textbf{each} $intent$ $\in$ $intents$}
            \State $retries \leftarrow max\_retries$
            \State $feedback \leftarrow \emptyset$
            \While{$retries > 0$}
                \State $params \leftarrow LLM_{parameters}(Conversation, Message, Tools[intent], Level, feedback)$
                \If{$Level, output \leftarrow Tools(intent, params, Level)$} \Comment{$output$ is a success or failure message.}
                    \State $outputs \leftarrow outputs \cup output$
                    \State \textbf{break}
                \Else
                    \State $feedback \leftarrow output$
                    \State $retries \leftarrow retries - 1$
                \EndIf
            \EndWhile
        \EndFor
        \State $response \leftarrow LLM_{summary}(outputs, Level)$
    \EndIf
    \State \Return $response$
\end{algorithmic}
\end{algorithm}

The proposed framework is intended to be a \enquote{drop-in} replacement for current tool-using pipelines. The end user would still see the entire process as a black box (\textit{i.e.}: their message is sent, some changes take place, and a message is produced), operating under acceptable time ranges\footnote{For the purposes of content design, a \enquote{real-time} response is in the 5 to 10 seconds range.}.

\section{Experimental Protocol}\label{sec:experimental_protocol}
In this work we assess the performance of the proposed framework, evaluating it on the existing \textit{LLMaker} test set \cite{gallotta2024consistent}. The test set is comprised of five test cases (T1 to T5) that mimic a designer's interaction with the tool. The test cases have an increasing number of requests and grow in complexity, with T5 being the hardest. The tool set is comprised of a total of 16 functions to add, edit, or remove rooms, corridors, enemies, treasures, traps, and enemy actions. As baseline comparison we run the same tests with the models that support tool usage via the Ollama API (henceforth, \enquote{Tools}). The JSON schema of the functions is approximately 3933 tokens as tokenized by GPT-2\footnote{We obtain this number by generating the schema following OpenAI's specifications (see \url{https://platform.openai.com/docs/assistants/tools/function-calling}. It is important to note that not all models share the same tokenizer, so the token count may differ slightly.}. We include in \Cref{tab:t5_example} the full test case 5 of the set of requests as an example.
\begin{table}[t!]
\centering
\caption{User requests for test case T5 as provided by \cite{gallotta2024consistent}. Each request is submitted sequentially.}
\label{tab:t5_example}
\begin{tabular}{|l|}
\hline
\rowcolor{lightgray!50} 
Create 3 rooms, each connected to the next one, all set in a different \\ 
\rowcolor{lightgray!50} 
European city                                                                         \\
\rowcolor{gray!50} 
Add a goblin archer in the first room                                                 \\
\rowcolor{lightgray!50} 
Also add two zombies                                                                  \\
\rowcolor{gray!50} 
Now generate a room connected to the first one, set in underground  \\
\rowcolor{gray!50} 
Atlantis                                                                              \\
\rowcolor{lightgray!50} 
Put a couple of evil mermaids in Atlantis                                             \\
\rowcolor{gray!50} 
Place multiple ocean-themed traps in the corridor to Atlantis                         \\
\rowcolor{lightgray!50} 
Place a single treasure chest in all rooms, each containing a piece of  \\
\rowcolor{lightgray!50} 
a treasure map                                                                        \\
\rowcolor{gray!50} 
Remove the chest containing the second piece of the treasure map                      \\
\rowcolor{lightgray!50} 
Add another room connected to Atlantis, set in Hell                                   \\
\rowcolor{gray!50} 
Place two fallen angels armed with flaming swords                                     \\
\rowcolor{lightgray!50} 
Change one of the angels to a capybara monster                                        \\
\rowcolor{gray!50} 
Set the health of the capybara to 1000                                                \\
\rowcolor{lightgray!50} 
Make the capybara a punker, with pink spiky hair                                      \\ \hline
\end{tabular}
\end{table}

In this work, we set to answer the following research questions:
\resarchq{How does \textsc{Freyr} perform compared to Tools?}
We are interested in comparing the performance of \textsc{Freyr} against Tools. For each of the test cases, we count the number of steps that lead to a successful edit (\textit{i.e.}: an edit that matches the designer's request). In their initial \textit{LLMaker} paper, Gallotta \textit{et al.} \cite{gallotta2024consistent} considered different LLM failure modes in level editing, including parser, domain, and design failures. As they were concerned with editing levels in a way that would match the designer's request, they would terminate the execution of a test case in case of a parser or domain failure, as any further change to the level would lead to an invalid output. We instead do not terminate the test case as soon as an error is generated. We execute each step of the test case, providing a predefined starting level for each step to properly assess the edits performed by the LLM. These hand-authored levels are always valid and ensure a consistent evaluation. After each step, we verify whether the LLM's changes satisfy both domain and design requirements, which are predefined based on Gallotta \textit{et al.} \cite{gallotta2024consistent}. A step is considered successful if it updates the level in a way that passes both domain and design validity checks. For example, for the second step in T5, \enquote{Add a goblin archer in the first room} (see \Cref{tab:t5_example}), the initial level for this step contains three empty rooms (one set in \enquote{Rome}, one in \enquote{Paris}, and one in \enquote{Barcelona}) and the step is considered successful if the only change is that the first room (the one in \enquote{Rome}) now contains an enemy.

\resarchq{Does \textsc{Freyr} require fewer tokens overall than Tools?}
LLMs operate by attending to tokens in input and generating a sequence of tokens as outputs. The input tokens generally consist of the system prompt provided to the LLM and the conversation messages. In Tools mode, the full description of each function is passed in JSON format in the system prompt, and additional commands are injected in the system prompt to condition the LLM to generate valid tool calls if needed. The number of input tokens then scales with the number of tools available to the LLM, which can lead to degraded performance. \textsc{Freyr}, instead, introduces an overhead because of multiple roles. Each role has a specific system prompt with its role's task definition. While this impacts the number of input tokens, each role also results in at least one output each, which affects the number of generated tokens. In this work we keep track of both the number of tokens in input ($\mathrm{Tokens}_\mathrm{in}$) and output ($\mathrm{Tokens}_\mathrm{out}$), as measured by the Ollama API, summing them across roles. Keeping the counts separate allows us to better understand where the overhead introduced by \textsc{Freyr} affects performance and gives a clearer indication of the scaling capabilities of either approach with a larger number of tools available.

\resarchq{Is \textsc{Freyr} slower than Tools?}
As introduced in \Cref{sec:freyr}, our framework assigns different roles to different LLMs that result in multiple responses being generated before a final output is returned by the system. Similarly, Tools also generates the tool calls before evaluating them. We are interested in comparing the overhead introduced by \textsc{Freyr} against Tools, as both systems suffer from a longer response time depending on the number of tool calls executed and their complexity. In this work we measure the cumulative time required to obtain a response from the system. For \textsc{Freyr} particularly, this time is the sum of time taken to generate a response by each role. For both modes, \textsc{Freyr} and Tools, we measure time in seconds even when a response results in an error. In case the error is generated internally, for example when a parameter for a function is not generated, both modes are instructed to try again and regenerate their response. The time it takes to regenerate a response until a valid one is produced is also tracked. If no valid response is generated after the maximum number of allowed retries, the test case step is marked as failed, but we still include that test case step's time in our results. In general, we want to obtain responses quickly to allow for real-time content editing. Here, we can also compare our results against the GPT-3.5 Turbo performance reported by Gallotta \textit{et al.} \cite{gallotta2024consistent}.

\subsection*{Models} We test the following LLMs: Llama 3.1 (8B), Qwen 2.5 (7B), Command-R (35B), and Gemma 2 (9B). For the \enquote{chat} and \enquote{summary} roles of \textsc{Freyr} we instead use Qwen 2.5 (0.5B). Each model has an input context length of 128k tokens. We set the temperature to 0.8 and nucleus sampling (\textit{top-p}) to 0.6. Interaction with these models is done via the Ollama API.

\subsection*{Experiment}
We execute all test cases from the \textit{LLMaker} test suite. We do not terminate on errors, and provide a starting predefined level for each step of each test case. We set the maximum number of allowed retries per request to three. We deploy \textsc{Freyr} with the same $LLM_{intent}$ and $LLM_{params}$ for a fair comparison against the Tools approach, as it uses a single model. All experiments were conducted on a single desktop computer equipped with an Nvidia RTX A5000 GPU, Intel Core i9-14900KF, and 128GB of RAM.

\subsection*{Baselines}
We compare performance of \textsc{Freyr} against tool usage as provided via the Ollama API (Tools). Not all models however are supported: the official versions of Gemma 2 are not among the listed supported models\footnote{\url{https://ollama.com/search?c=tools}} at the time of writing and were thus discarded for comparison. We analyze however the performance for all models combinations in \textsc{Freyr} as an ablation study (in \Cref{subsec:freyr_models_ablstudy}).

\section{Results}\label{sec:results}
In this section we report the results of \textsc{Freyr} and tool usage via Ollama API (Tools) approaches on the \textit{LLMaker} test set (\Cref{subsec:freyr_vs_tools}). We then analyze the effects of models choice in the different blocks of \textsc{Freyr} (\Cref{subsec:freyr_models_ablstudy}). Finally, we comment on the quality of the generated content from a designer perspective (\Cref{subsec:creativity_results}).

\subsection{Performance of \textsc{Freyr} against Tools}\label{subsec:freyr_vs_tools}
We evaluated the performance of \textsc{Freyr} and compared it against Tools on the \textit{LLMaker} test set. We present our results in \Cref{tab:freyr_tools_single}. A Wilcoxon signed-rank test with Bonferroni correction ($p<0.05$) was conducted to determine statistical difference between results. 

\begin{table}[!th]
\caption{Performance for \textsc{Freyr} and Ollama Tools approaches for all 5 test cases of \textit{LLMaker} \cite{gallotta2024consistent}. We report the percentage of successful steps completed, the cumulative input and output tokens (including retries) and execution time per step. Reported values are averaged over 10 runs with 95\% credible intervals. Best values per test case on each metric are in \textbf{bold}, with ${}^\dagger$ indicating statistical significance between modes.}
\label{tab:freyr_tools_single}
\centering
\begin{tabular}{|cllllll|}
\hline
\multicolumn{1}{|c|}{Test Case} & \multicolumn{1}{l|}{LLM} & \multicolumn{1}{l|}{Mode} & \multicolumn{1}{l|}{Steps (\%) ($\uparrow$)} & \multicolumn{1}{l|}{Tokens${}_\mathrm{in}$ ($\downarrow$)} & \multicolumn{1}{l|}{Tokens${}_\mathrm{out}$ ($\downarrow$)} & Time (s) ($\downarrow$) \\ \hline
\multirow{6}{*}{T1}              & \multirow{2}{*}{Command-R} & \textsc{Freyr}           & $83 \pm 4$                   				& $2816 \pm 195$         			   & $213 \pm 11$    								& $8.5 \pm 0.5$   \\
                                &                            & Tool                      & $47 \pm 4$                   				& $6930 \pm 1002$        			   & $125 \pm 39$    								& $52.0 \pm 13.6$ \\ \cline{3-7} 
                                & \multirow{2}{*}{Llama 3.1} & \textsc{Freyr}            & $\mathbf{86 \pm 0^\dagger}$          				& $2545 \pm 62$          			   & $226 \pm 33$    								& $3.2 \pm 0.3$   \\
                                &                            & Tool                      & $41 \pm 7$                   				& $4586 \pm 273$         			   & $181 \pm 17$    								& $4.7 \pm 0.4$   \\ \cline{3-7} 
                                & \multirow{2}{*}{Qwen 2.5}  & \textsc{Freyr}            & $71 \pm 0$                   				& $\mathbf{2041 \pm 84}^\dagger$ 			   & $\mathbf{166 \pm 14}^\dagger$    						& $\mathbf{2.5 \pm 0.1}^\dagger$   \\
                                &                            & Tool                      & $37 \pm 6$                   				& $9868 \pm 710$         			   & $316 \pm 24$    								& $7.0 \pm 0.5$   \\ \hline \hline
\multirow{6}{*}{T2}              & \multirow{2}{*}{Command-R} & \textsc{Freyr}           & $\mathbf{98 \pm 3^\dagger}$                   		& $2242 \pm 65$          			   & $141 \pm 4$     								& $6.0 \pm 0.2$   \\
                                &                            & Tool                      & $27 \pm 7$                   				& $7787 \pm 498$         			   & $123 \pm 13$    								& $52.9 \pm 4.5$  \\ \cline{3-7} 
                                & \multirow{2}{*}{Llama 3.1} & \textsc{Freyr}            & $71 \pm 4$                   				& $2811 \pm 164$         			   & $274 \pm 11$    								& $3.9 \pm 0.1$   \\
                                &                            & Tool                      & $38 \pm 8$                   				& $4542 \pm 135$         			   & $156 \pm 31$    								& $4.4 \pm 0.3$   \\ \cline{3-7} 
                                & \multirow{2}{*}{Qwen 2.5}  & \textsc{Freyr}            & $89 \pm 0$          				            & $\mathbf{1582 \pm 1^\dagger}$  			   & $\mathbf{90 \pm 1}^\dagger$      				& $\mathbf{1.6 \pm 0.0}^\dagger$   \\
                                &                            & Tool                      & $21 \pm 5$                   				& $7765 \pm 883$         			   & $188 \pm 19$    						& $5.1 \pm 0.5$   \\ \hline \hline
\multirow{6}{*}{T3}              & \multirow{2}{*}{Command-R} & \textsc{Freyr}           & $\mathbf{90 \pm 5^\dagger}$          				& $2961 \pm 123$         			   & $193 \pm 7$     								& $8.2 \pm 0.3$   \\
                                &                            & Tool                      & $36 \pm 6$                   				& $8150 \pm 478$         			   & $112 \pm 20$    								& $50.3 \pm 7.4$  \\ \cline{3-7} 
                                & \multirow{2}{*}{Llama 3.1} & \textsc{Freyr}            & $60 \pm 4$                   				& $2695 \pm 226$         			   & $318 \pm 33$    						& $4.3 \pm 0.4$   \\
                                &                            & Tool                      & $28 \pm 6$                   				& $4721 \pm 153$         			   & $142 \pm 22$    						& $4.5 \pm 0.3$   \\ \cline{3-7} 
                                & \multirow{2}{*}{Qwen 2.5}  & \textsc{Freyr}            & $86 \pm 3$          				& $\mathbf{1906 \pm 21}^\dagger$ 			   & $\mathbf{93 \pm 3}^\dagger$      				& $\mathbf{1.7 \pm 0.0^\dagger}$   \\
                                &                            & Tool                      & $23 \pm 6$                   				& $8116 \pm 1089$        			   & $180 \pm 32$    						& $5.3 \pm 0.8$   \\ \hline \hline
\multirow{6}{*}{T4}              & \multirow{2}{*}{Command-R} & \textsc{Freyr}           & $\mathbf{68 \pm 4^\dagger}$          				& $2717 \pm 178$         			   & $205 \pm 20$    								& $8.4 \pm 0.7$   \\
                                &                            & Tool                      & $25 \pm 3$                   				& $6896 \pm 421$         			   & $\mathbf{97 \pm 20}^\dagger$     						& $44.0 \pm 7.1$  \\ \cline{3-7} 
                                & \multirow{2}{*}{Llama 3.1} & \textsc{Freyr}            & $54 \pm 2$                   				& $2966 \pm 234$         			   & $233 \pm 34$    								& $3.5 \pm 0.4$   \\
                                &                            & Tool                      & $17 \pm 3$                   				& $4868 \pm 220$         			   & $175 \pm 32$    								& $5.0 \pm 0.4$   \\ \cline{3-7} 
                                & \multirow{2}{*}{Qwen 2.5}  & \textsc{Freyr}            & $52 \pm 3$                   				& $\mathbf{1793 \pm 78}^\dagger$ 			   & $118 \pm 10$    								& $\mathbf{1.9 \pm 0.1}^\dagger$   \\
                                &                            & Tool                      & $25 \pm 4$                   				& $5634 \pm 398$         			   & $132 \pm 15$    								& $3.9 \pm 0.3$   \\ \hline \hline
\multirow{6}{*}{T5}              & \multirow{2}{*}{Command-R} & \textsc{Freyr}           & $\mathbf{51 \pm 5^\dagger}$          				& $4771 \pm 112$         			   & $252 \pm 7$     								& $10.9 \pm 0.3$  \\
                                &                            & Tool                      & $28 \pm 9$                   				& $8202 \pm 870$         			   & $\mathbf{226 \pm 72}$    						& $95.0 \pm 29.2$ \\ \cline{3-7} 
                                & \multirow{2}{*}{Llama 3.1} & \textsc{Freyr}            & $12 \pm 3$                   				& $3679 \pm 228$                 			   & $1077 \pm 1242$ 								& $8.1 \pm 4.6$   \\
                                &                            & Tool                      & $29 \pm 5$                   				& $7492 \pm 852$                 			   & $417 \pm 99$    								& $9.0 \pm 1.4$   \\ \cline{3-7} 
                                & \multirow{2}{*}{Qwen 2.5}  & \textsc{Freyr}            & $28 \pm 2$                   				& $\mathbf{3004 \pm 194}^\dagger$			   & $794 \pm 1229$  								& $\mathbf{4.9 \pm 4.4}^\dagger$   \\
                                &                            & Tool                      & $44 \pm 6$                   				& $9072 \pm 1338$        			   & $335 \pm 77$    								& $7.8 \pm 1.3$   \\ \hline
\end{tabular}
\end{table}

\finding{\textsc{Freyr} consistently outperforms Tools in handling requests.}

When focusing on the number of steps completed successfully per test case (Steps (\%)), we can see that \textsc{Freyr} outperforms its counterpart in Tools in 13 out of 15 cases, completing more steps. The two cases where \textsc{Freyr} underperforms are with Llama 3.1 and Qwen 2.5 on T5, which is the most difficult test case. We can see that, regardless of the model, the Tools approach only achieves at most 47\% and as low as 17\% of completed steps. \textsc{Freyr} can almost (98\%) complete entire test cases (T2) or reach 80\% to 90\% completion rates (T1 and T3). The generally best performing model is Command-R, which is also the largest model. Except in T1, where Llama 3.1 scores higher, Command-R dominates in all other test cases, with Qwen 2.5 close behind (T2 and T3). However, understanding failures is of equal importance as highlighting success. Reading through the results logs, we find that most of the Tools failures are due to repeated wrong tools parameters being generated. This type of failure is handled by querying the LLM to regenerate the response, providing it with details on the previous error. While we found that sometimes errors would be fixed in one or two retries, most instances would fail to solve the error, resulting in a failure due to exceeding the maximum number of retries allowed. For \textsc{Freyr}, instead, we highlight the failure case of Llama 3.1 in T5, which is the only case where Tools outperforms \textsc{Freyr}. Here, the model fails by generating too many intents, which triggers an early termination. The problem of token repetitions in LLM responses is well known \cite{zhang2023joint}, and can generally be avoided with careful parameter tuning. 

\finding{\textsc{Freyr} consumes less tokens than Tools, but tends to generate more tokens.}
When looking at tokens, we find that \textsc{Freyr} uses fewer tokens in input ($\mathrm{Tokens}_{\mathrm{in}}$) than Tools, ranging from 57\% of the tokens in the case of Llama 3.1 in T3 to 20\% in the case of Qwen 2.5 in T3. This trend is consistently found across all models tested on all test cases. Qwen 2.5 notably uses the fewest tokens in input on all test cases. For the output tokens ($\mathrm{Tokens}_{\mathrm{out}}$), results show a different trend. While Qwen 2.5 in \textsc{Freyr} generates fewer output tokens than its Tools counterpart on four out of five test cases, this difference is significant only in two test cases (T2 and T3). Command-R and Llama 3.1 tend to generate more output tokens with \textsc{Freyr} than with Tools, although the differences are not significant. Command-R also generates the fewest output tokens in Tools mode for T4 and T5, though again not significantly. We will revisit whether this behavior leads to better responses in \Cref{subsec:creativity_results}. We highlight two particularities found in T5, the hardest test case. First, Qwen 2.5 in \textsc{Freyr} generates more than double the output tokens of its Tools counterpart, which is in stark contrast with its performance on the other test cases. Secondly, Llama 3.1 in \textsc{Freyr} also generates more than double the output tokens than its Tools counterpart. In both cases we note that the results are extremely noisy, likely due to the random initialization of the models. However, this should not be a surprise: as mentioned above, the failure of Llama 3.1 on this test case is due to the generation of too many intents, which impact the number of tokens generated. For Qwen 2.5, instead, we find from the results logs that this higher-than-usual number of output tokens is due to a failure of self-correcting ill-formed tool calls. This also explains the fewer successful responses.

\finding{\textsc{Freyr} has consistently and significantly lower response times than Tools.}
Finally, we can look at the average step time elapsed across test cases. As the speed at which a step is completed is tightly related to the speed at which prompts are processed and responses are generated, it comes at no surprise that the models with fewer input and output tokens also achieve faster response times. We find that Qwen 2.5 in \textsc{Freyr} is significantly faster than its Tools counterpart, with elapsed time savings that range from as little as 52\% on T5 to as much as 70\% on T3. Llama 3.1 instead achieves comparable times in either modes, with a significant difference only on T1 and T4. The most interesting outcome of these results is however Command-R. We find that this model in \textsc{Freyr} consistently outperforms its Tools counterpart, with times ranging from approximately 11\% (T2 and T5) to 16\% (T1, T3, and T4) of the elapsed time in Tools. These time reductions are more impressive than the ones for any other model tested. However, we note that the actual time taken by Command-R in \textsc{Freyr} mode may be too much for a real-time application, as it is too close to the upper bound of 10 seconds. Its Tools counterpart is simply unusable in real-time applications, as it can take up to a minute before a step is completed. Qwen 2.5 and Llama 3.1 in \textsc{Freyr} instead generate responses in between 2 to 5 seconds for Qwen2.5, and between 3 to 8 seconds for Llama 3.1. Their percentage of steps completed exceeds their Tools counterparts in all test case except T5. While completing less steps than Command-R in \textsc{Freyr}, the trade-off between speed and accuracy make both of these two models great candidates for real-time applications

We can also compare \textsc{Freyr} with GPT-3.5 Turbo, based on the results reported by Gallotta \textit{et al.} \cite{gallotta2024consistent}. In their paper, they report that tool usage with GPT-3.5 Turbo consistently solved all test cases with no failures. Interestingly, the reported time per request on each test case ranges from approximately 5 to 10 seconds, which are in line with the time taken by the open-source models in Tools mode. However, with \textsc{Freyr}, we achieve lower average response times than GPT-3.5 Turbo, though we do not achieve the same completion rates.

\subsection{Models Sensitivity in \textsc{Freyr}}\label{subsec:freyr_models_ablstudy}
One of the key strengths in \textsc{Freyr} is its separate modules, that allow for easy drop-in replacement and combinations of LLMs to take place. While in the previous section we analyzed the performance of \textsc{Freyr} when using the same LLM for both intent recognition and parameters generation steps, here we instead focus on the effects that different models have when playing different roles. We also evaluate the performance of \textsc{Freyr} using a model that is not supported directly for tool usage via the Ollama API. Here, we use Gemma 2 (7B). We include the full table with results in \Cref{app:ablstudy}.

\finding{The choice of different LLMs for different roles does not significantly affect performance in \textsc{Freyr}.}
We find that employing separate models for intent detection ($LLM_{intent}$) and parameter generation ($LLM_{parameters}$) results in a modest performance improvement compared to using the same model for both roles. For instance, Qwen 2.5 alone achieves only 89\% of completed steps on T2, but it successfully completes the test case when paired with either Command-R or Gemma 2 as $LLM_{parameters}$. Similarly, Llama 3.1's performance on T3 increases dramatically from 60\% to 99\% when Gemma 2 is used as $LLM_{parameters}$. However, not all configurations lead to improvements: in T4, Qwen 2.5's performance drops from 52\% when handling both roles to 51\% when paired with Command-R as $LLM_{parameters}$. Despite occasional drops, performance gains outweigh losses overall. The largest improvement is a 33\% increase by Llama 3.1 as $LLM_{intent}$ and Gemma 2 as $LLM_{parameters}$ on T3, while the largest decrease is a 17\% drop by Command-R as $LLM_{intent}$ and Llama 3.1 as $LLM_{parameters}$ on T2.

When focusing on configurations where Gemma 2 serves as $LLM_{parameters}$, we find that these pairings achieve some of the highest percentages of completed steps across test cases, often nearing or reaching 100\% (86\% on T1, 100\% on T2, 100\% on T3, 73\% on T4, and 55\% on T5). However, some combinations are slower to generate responses. For example, in T3 and T5, Qwen 2.5 as $LLM_{intent}$ is the second slowest configuration (13.7 seconds and 15.1 seconds for T3 and T5, respectively). Gemma 2 also performs exceptionally well when paired with itself for both $LLM_{intent}$ and $LLM_{parameters}$, consistently achieving near-perfect completion rates (e.g., 100\% in T3 and T5). Additionally, this configuration is faster than Qwen 2.5, making it a strong choice for time-sensitive tasks without compromising accuracy.

That said, Gemma 2's standalone performance shows variability. For example, on T2, it completes 11\% fewer steps than the best-performing configuration, which uses Qwen 2.5 as $LLM_{intent}$ and Gemma 2 as $LLM_{parameters}$. Similarly, on T5, it completes 17\% fewer steps compared to the best configuration, which pairs Command-R as $LLM_{intent}$ with Gemma 2 as $LLM_{parameters}$. These results highlight trade-offs between performance and efficiency. For instance, using Llama 3.1 as $LLM_{intent}$ with Qwen 2.5 as $LLM_{parameters}$ achieves the fastest response times in four out of five test cases (except T5), but ranks last in terms of completed steps.

The combination of Gemma 2 as $LLM_{intent}$ and Llama 3.1 as $LLM_{parameters}$ resulted in the lowest input token usage in two out of five test cases (T1 and T5). Notably, the configuration that processes the fewest input tokens also tends to generate the fewest output tokens across test cases. This trend differs from the results shown in \Cref{tab:freyr_tools_single}, where, for instance, in T5, Command-R generates 252 tokens compared to Qwen 2.5's 794 tokens, despite attending to nearly 2,000 more tokens. Additionally, certain model combinations further optimized token usage. For example, in T4, the pairing of Llama 3.1 as $LLM_{intent}$ and Command-R  as $LLM_{parameters}$ processed only 60\% and 55\% of the input tokens, respectively, and generated just 39\% and 45\% of the output tokens compared to their standalone performance.

Overall, the results indicate that while there are exceptions and nuances based on specific configurations and test cases, the choice of different LLMs for different roles can lead to minor improvements. As there is no specific combination that performs better across all metrics on all test cases, the configuration must be selected in an ad-hoc fashion based on specific performance criteria.

\subsection{Creativity: Beyond Validity}\label{subsec:creativity_results}
So far we have been concerned with the response and the performance of the framework. In \textit{LLMaker}, however, the response provided by the system (generated by $LLM_{summary}$ in \textsc{Freyr}) is only half the feedback to the designer. The application generates \textit{content} as well, setting its properties that ultimately drive the creative process. One interesting aspect of the generated content that we did not evaluate objectively is its \emph{quality}. If the model produces uninteresting content, it would lead to a less appealing interaction for the user. If instead the model generates serendipitous content, the user would be more prone to explore different design choices. There are many and different measures of creativity, each with its own limitations \cite{franceschelli2024creativity}. Here however we just aim to give a qualitative evaluation, according to the authors' judgment. \Cref{tab:creativity_examples} reports two example contents edited during a test case step per model sampled at random from all runs and test cases.

\begin{table}[t!]
\centering
\caption{Randomly sampled level edits for both \textsc{Freyr} and Tools. A level \enquote{edit} is a change made to the level driven by the LLM. We omit properties that do not play a role in the creative evaluation of a given content type (for example, we omit the health points of enemies or the direction a new room is connected to another).}
\label{tab:creativity_examples}
\begin{tabular}{|p{0.1\linewidth}p{0.25\linewidth}p{0.05\linewidth}p{0.55\linewidth}|}
\hline
\multicolumn{1}{|l|}{LLM}  & \multicolumn{1}{l|}{Query}                                          & \multicolumn{1}{l|}{Mode} & Edit Details ("Name": "Description")                                                                                                                                                                   \\ \hline
\multirow{4}{*}{Command-R} & Create a room with a gravity-defying effect where everything floats & \textsc{Freyr}                     & "The Floating Chamber": "A spacious chamber with an ethereal atmosphere, featuring zero-gravity effects that allow objects and occupants to float freely."                                             \\ \cline{3-4} 
                           &                                                                     & Tools                     & "Floating Room": "A room with a gravity-defying effect where everything floats."                                                                                                                       \\ \cline{2-4} 
                           & Make a new room filled with ancient runes on the floor              & \textsc{Freyr}                     & "Ancient Runes Chamber": "A dimly lit chamber with intricate ancient runes covering the floor. The air is thick with an otherworldly energy, and the runes seem to pulse softly with an unseen power." \\ \cline{3-4} 
                           &                                                                     & Tools                     & "ancient\_runes\_room": "filled with ancient runes on the floor"                                                                                                                                       \\ \hline
\multirow{4}{*}{Llama 3.1} & \multirow{2}{*}{Add a couple of enemies}                            & \textsc{Freyr}                     & "SwampySnake": "A venomous snake with mottled brown skin"; "Enemy SwampySnake": "A venomous snake with mottled brown skin"                                                                             \\ \cline{3-4} 
                           &                                                                     & Tools                     & "Gob 1": "A small, nimble goblin."; "Gob 2": A slightly larger, scarier goblin."                                                                                                                       \\ \cline{2-4} 
                           & \multirow{2}{*}{Place a treasure chest}                             & \textsc{Freyr}                     & "Golden Idol": "A small, intricately carved golden idol with a gemstone in its forehead."                                                                                                              \\ \cline{3-4} 
                           &                                                                     & Tools                     & "treasure chest": "a small wooden chest"                                                                                                                                                               \\ \hline
\multirow{4}{*}{Qwen 2.5}  & Create a room with a collapsed bridge spanning a dark chasm         & \textsc{Freyr}                     & "Room with Collapsed Bridge": "A narrow passage with a partially collapsed stone bridge, the remnants of which hang precariously over a deep, shadowy chasm."                                          \\ \cline{3-4} 
                           &                                                                     & Tools                     & "BridgeRoom": "A room with a collapsed bridge spanning a dark chasm. The air is thick with the smell of decay and the distant echo of dripping water."                                                 \\ \cline{2-4} 
                           & Add a goblin archer in the first room                               & \textsc{Freyr}                     & "Goblin Archer": "A small, agile humanoid with a bow and quiver full of arrows."                                                                                                                       \\ \cline{3-4} 
                           &                                                                     & Tools                     & "GoblinArcher": "A nimble and quick archer, equipped with a bow."                                                                                                                                      \\ \hline
\end{tabular}%
\end{table}

\finding{Content generated via \textsc{Freyr} is more creative than that generated via Tools.}

Based on \Cref{tab:creativity_examples}, we can see that Tools seem to reuse heavily any detail introduced in the query: for example, Command-R in Tools sets the description of the new room as \enquote{filled with ancient runes on the floor} verbatim, whereas the \textsc{Freyr} version of the same model generates a more complex description while adhering to the description provided in the query. Similarly, we find that Tools simply gives numbers to differentiate content of the same type (\enquote{Gob 1} and \enquote{Gob 2}), while \textsc{Freyr} occasionally fails to generate fully distinct content within the same type (\enquote{SwampySnake} and \enquote{Enemy SwampySnake}). These failures often require users to intervene manually. For instance, the user may request to rename entities to avoid ambiguity. Such interruptions can slow down the design process and risk frustrating the user, particularly if these issues occur frequently.

Nonetheless, we find that \textsc{Freyr} generates more creative content. We believe that this creativity is is able to emerge because the generation of parameters in \textsc{Freyr} occurs in its separate step, which lets the LLM focus on generating more freely. With Tools, instead, this generation occurs in a more restrictive frame, as the LLM is requested to produce a valid JSON as response.

\section{Discussion}
From the presented results, we find that our proposed framework, \textsc{Freyr}, consistently outperforms the existing approach to enable LLMs to use tools via the Ollama API. We explored the impact on performance of assigning tasks to different LLMs in our framework, and demonstrated how \textsc{Freyr} can enable LLMs for tool usage even if they do not natively support it, with no degradation of quality. While not explored, the framework can be further tuned to different needs by tinkering with the prompts assigned to the different roles (\textit{e.g.}: $LLM_{intent}$). For example, if explainability is of interest, $LLM_{intent}$ can be set to also produce an explanation for each detected intent. This level of information however would not be as informative as an entire thought process as available with more recent finetuned LLMs \cite{nexusraven}. While we developed \textsc{Freyr} with open-source and full control of the flow of information in mind, the framework can be applied to closed-source and proprietary models (such as OpenAI's GPT-4 or Anthropic's Claude 3.5), as it relies only on prompt engineering techniques. The lack of control, replicability, and environmental impact of these massive models is however a concern \cite{gallotta2024llmandgames}, regardless of how well they might perform on these tasks.

This work has multiple limitations: mainly, we evaluated \textsc{Freyr} only on a single domain. While the \textit{LLMaker} test cases proved challenging for existing methods already, it does not provide any guarantee on generalizability of results. Secondly, we only evaluate a small subset of all possible existing LLMs, focusing on models with medium context lengths. While long-context LLMs would probably be better suited when many tools are provided, they are also known to struggle with in-context learning tasks \cite{li2024longcontext}. However, more recent models can handle longer context without degradation in performance \cite{li2024longcontext, xiao2024efficient}. Another limitation we highlight in this work is that \textsc{Freyr} enforces a simple yet structured format in the responses (comma-separated list for intents, and bulleted list for the parameters). The appeal of instead enforcing more structured outputs, such as JSON or XML, in the generation of responses by the LLM is sought after by industry professionals as it would speed up prompt-based development efficiency, satisfy requirements, and improve user experience \cite{liu2024structured}. There are multiple approaches that allow for such structured generation via controlled grammar decoding \cite{ugare2024syncodellm, geng2023grammar, park2024gad}, where the schema of the response is defined by the user and enforce in the response. In preliminary tests for this paper using the Outlines library \cite{willard2023efficient}, however, we found that such approaches could lead to unstable responses and sometimes would cause the entire generation infinitely repeat the same sequence of tokens.

One last limitation of the presented work is that we do not focus on the quality of the generated content. While we do give a qualitative overview of the results in \Cref{subsec:creativity_results}, we do not formally evaluate it, either automatically or via human evaluations. Creativity in LLMs is still a hot topic in the computational creativity community \cite{franceschelli2024onthecreativity}, and we believe \textsc{Freyr} could benefit from a user study. In concordance to existing literature in the field, \textsc{Freyr} generates seemingly creative content by setting a non-zero temperature \cite{peeperkorn2024temperature}, not requesting the LLM to \enquote{just be creative} \cite{veal2024symbolic}, and using prompts that allow for creative writing \cite{bellemarepepin2024divergent}. However, LLMs are known to only be capable of valuablee creativity and a weak version of novelty, possibly due to their autoregressive nature \cite{franceschelli2024creativity}.

\section{Conclusions}
In this work, we introduced \textsc{Freyr}, a novel framework designed to allow tool usage with any LLM by modularizing the tool usage process. Through comprehensive experimentation on the real-world test case specific to video game design, \textit{LLMaker}, we demonstrated that \textsc{Freyr} consistently outperforms the traditional approach via the Ollama API. By decomposing the tool usage process into distinct steps, \textsc{Freyr} enables LLMs to effectively utilize tools without requiring model-specific adaptations or fine-tuning, thereby overcoming key limitations of existing methodologies. By also releasing the code for this work publicly, we hope that \textsc{Freyr} will inspire more research in the area of flexible, innovative, and effective tool usage for LLMs.

\section*{Ethics Statement}
This work relies on pre-trained language models, which are trained on large corpora of text extracted from the web and are known to contain biased and discriminatory content \cite{gallotta2024llmandgames}. As the presented work relies on the knowledge of the underlying models, content generated by this system may present the same type of biases. We emphasize the importance of careful curation and supervision by human users to ensure the generated content aligns with ethical standards and avoids perpetuating stereotypes and harmful narratives. This also includes clear communication to users about AI-generated content and its potential limitations.

We also recognize the environmental impact of deploying large-scale language models. Efforts were made to use computational resources responsibly, and we encourage future research to explore more energy-efficient methodologies.

\section*{Acknowledgments}
The authors would like to thank Matthew Barthet and Nemanja Rasajski for the insightful discussions over Friday's lunches.

\bibliographystyle{plainnat}
\bibliography{references}

\break

\appendix

\section{Other Models Combinations in \textsc{Freyr}}\label{app:ablstudy}
In this section we present the complete results of the tested models (Command-R, Gemma 2, Llama 3.1, and Qwen 2.5) in \textsc{Freyr}.

\begin{longtable}{|cllllll|}
\caption{Performance for missing \textsc{Freyr} configurations on all 5 test cases. We report the percentage of successful steps completed, the cumulative input and output tokens (including retries) and execution time per step. Reported values are averaged over 10 runs with 95\% credible intervals. Best values per test case on each metric are in \textbf{bold}, with ${}^\dagger$ indicating statistical significance among configurations.}
\label{tab:freyr_single_pt1}\\
\hline
\multicolumn{1}{|c|}{Test Case} & \multicolumn{1}{l|}{$LLM_{intent}$} & \multicolumn{1}{l|}{$LLM_{parameters}$} & \multicolumn{1}{l|}{Steps (\%) ($\uparrow$)} & \multicolumn{1}{l|}{Tokens${}_\mathrm{in}$ ($\downarrow$)} & \multicolumn{1}{l|}{Tokens${}_\mathrm{out}$ ($\downarrow$)} & Time (s) ($\downarrow$)        \\ \hline
\endfirsthead
\multicolumn{7}{c}%
{{\bfseries Table \thetable\ continued from previous page}} \\
\hline
\multicolumn{1}{|c|}{Test Case} & \multicolumn{1}{l|}{$LLM_{intent}$} & \multicolumn{1}{l|}{$LLM_{parameters}$} & \multicolumn{1}{l|}{Steps (\%) ($\uparrow$)} & \multicolumn{1}{l|}{Tokens${}_\mathrm{in}$ ($\downarrow$)} & \multicolumn{1}{l|}{Tokens${}_\mathrm{out}$ ($\downarrow$)} & Time (s) ($\downarrow$)        \\ \hline
\endhead
\multirow{13}{*}{T1}            & \multirow{3}{*}{Command-R}          & Gemma 2                                 & $81\pm4$                                     & $2458 \pm 191$                                             & $178 \pm 9$                                                 & $12.2 \pm 0.4$                 \\ \cline{3-7} 
                                &                                     & Llama 3.1                               & $\mathbf{86\pm0}$                            & $2286 \pm 145$                                             & $204 \pm 12$                                                & $8.1 \pm 0.2$                  \\ \cline{3-7} 
                                &                                     & Qwen 2.5                                & $80\pm5$                                     & $2842 \pm 187$                                             & $217 \pm 17$                                                & $8.5 \pm 0.2$                  \\ \cline{2-7} 
                                & \multirow{4}{*}{Gemma 2}            & Command-R                               & $84\pm3$                                     & $2609 \pm 32$                                              & $188 \pm 8$                                                 & $13.0 \pm 0.3$                 \\ \cline{3-7} 
                                &                                     & Gemma 2                                 & $\mathbf{86\pm0}$                            & $2178 \pm 42$                                              & $154 \pm 5$                                                 & $5.7 \pm 0.2$                  \\ \cline{3-7} 
                                &                                     & Llama 3.1                               & $\mathbf{86\pm0}$                            & $2214 \pm 94$                                              & $186 \pm 8$                                                 & $3.1 \pm 0.2$                  \\ \cline{3-7} 
                                &                                     & Qwen 2.5                                & $\mathbf{86\pm0}$                            & $2442 \pm 87$                                              & $177 \pm 8$                                                 & $6.5 \pm 0.1$                  \\ \cline{2-7} 
                                & \multirow{3}{*}{Llama 3.1}          & Command-R                               & $84\pm3$                                     & $2424 \pm 48$                                              & $168 \pm 7$                                                 & $10.2 \pm 0.2$                 \\ \cline{3-7} 
                                &                                     & Gemma 2                                 & $80\pm5$                                     & $\mathbf{2058 \pm 32}$                                     & $\mathbf{132 \pm 10}^\dagger$                               & $8.0 \pm 0.2$                  \\ \cline{3-7} 
                                &                                     & Qwen 2.5                                & $80\pm5$                                     & $2280 \pm 20$                                              & $162 \pm 12$                                                & $\mathbf{2.4 \pm 0.1}^\dagger$ \\ \cline{2-7} 
                                & \multirow{3}{*}{Qwen 2.5}           & Command-R                               & $71\pm0$                                     & $3104 \pm 99$                                              & $271 \pm 32$                                                & $13.1 \pm 0.7$                 \\ \cline{3-7} 
                                &                                     & Gemma 2                                 & $71\pm0$                                     & $2513 \pm 74$                                              & $220 \pm 22$                                                & $10.0 \pm 0.4$                 \\ \cline{3-7} 
                                &                                     & Llama 3.1                               & $71\pm0$                                     & $2546 \pm 159$                                             & $238 \pm 14$                                                & $3.4 \pm 0.2$                  \\ \hline

\multirow{13}{*}{T2}            & \multirow{3}{*}{Command-R}          & Gemma 2                                 & $99\pm2$                                     & $2078 \pm 35$                                              & $130 \pm 3$                                                 & $10.6 \pm 0.2$                 \\ \cline{3-7} 
                                &                                     & Llama 3.1                               & $81\pm3$                                     & $2029 \pm 95$                                              & $146 \pm 8$                                                 & $7.4 \pm 0.2$                  \\ \cline{3-7} 
                                &                                     & Qwen 2.5                                & $83\pm4$                                     & $2406 \pm 136$                                             & $148 \pm 9$                                                 & $7.7 \pm 0.2$                  \\ \cline{2-7} 
                                & \multirow{4}{*}{Gemma 2}            & Command-R                               & $91\pm3$                                     & $2040 \pm 221$                                             & $120 \pm 18$                                                & $10.6 \pm 0.7$                 \\ \cline{3-7} 
                                &                                     & Gemma 2                                 & $89\pm3$                                     & $2228 \pm 153$                                             & $124 \pm 9$                                                 & $5.0 \pm 0.3$                  \\ \cline{3-7} 
                                &                                     & Llama 3.1                               & $84\pm4$                                     & $\mathbf{1659 \pm 70}$                                     & $\mathbf{97 \pm 8}$                                         & $2.0 \pm 0.1$                  \\ \cline{3-7} 
                                &                                     & Qwen 2.5                                & $84\pm4$                                     & $2265 \pm 216.3$                                           & $132 \pm 17$                                                & $6.0 \pm 0.3$                  \\ \cline{2-7} 
                                & \multirow{3}{*}{Llama 3.1}          & Command-R                               & $81\pm3$                                     & $1761 \pm 43$                                              & $100 \pm 4$                                                 & $8.1 \pm 0.1$                  \\ \cline{3-7} 
                                &                                     & Gemma 2                                 & $86\pm7$                                     & $1831 \pm 47$                                              & $104 \pm 2$                                                 & $7.4 \pm 0.1$                  \\ \cline{3-7} 
                                &                                     & Qwen 2.5                                & $71\pm7$                                     & $1881 \pm 59$                                              & $103 \pm 3$                                                 & $\mathbf{1.7 \pm 0.0}$         \\ \cline{2-7} 
                                & \multirow{3}{*}{Qwen 2.5}           & Command-R                               & $\mathbf{100\pm0}$                           & $2997 \pm 186$                                             & $295 \pm 25$                                                & $12.2 \pm 0.6$                 \\ \cline{3-7} 
                                &                                     & Gemma 2                                 & $\mathbf{100\pm0}$                           & $2702 \pm 133$                                             & $264 \pm 21$                                                & $10.0 \pm 0.4$                 \\ \cline{3-7} 
                                &                                     & Llama 3.1                               & $89\pm0$                                     & $2389 \pm 234$                                             & $267 \pm 28$                                                & $3.6 \pm 0.3$                  \\ \hline

\multirow{13}{*}{T3}            & \multirow{3}{*}{Command-R}          & Gemma 2                                 & $94\pm3$                                     & $2907 \pm 134$                                             & $169 \pm 7$                                                 & $12.1 \pm 0.3$                 \\ \cline{3-7} 
                                &                                     & Llama 3.1                               & $73\pm4$                                     & $3171 \pm 197$                                             & $209 \pm 16$                                                & $8.4 \pm 0.3$                  \\ \cline{3-7} 
                                &                                     & Qwen 2.5                                & $85\pm3$                                     & $2766 \pm 164$                                             & $175 \pm 14$                                                & $8.0 \pm 0.3$                  \\ \cline{2-7} 
                                & \multirow{4}{*}{Gemma 2}            & Command-R                               & $99\pm2$                                     & $2182 \pm 99$                                              & $116 \pm 9$                                                 & $10.7 \pm 0.3$                 \\ \cline{3-7} 
                                &                                     & Gemma 2                                 & $\mathbf{100\pm0}$                           & $2427 \pm 51$                                              & $126 \pm 2$                                                 & $5.2 \pm 0.1$                  \\ \cline{3-7} 
                                &                                     & Llama 3.1                               & $93\pm3$                                     & $2489 \pm 52$                                              & $143 \pm 7$                                                 & $2.8 \pm 0.1$                  \\ \cline{3-7} 
                                &                                     & Qwen 2.5                                & $87\pm3$                                     & $2288 \pm 90$                                              & $121 \pm 8$                                                 & $5.9 \pm 0.2$                  \\ \cline{2-7} 
                                & \multirow{3}{*}{Llama 3.1}          & Command-R                               & $93\pm4$                                     & $\mathbf{1880 \pm 47}^\dagger$                             & $\mathbf{92 \pm 2}$                                         & $8.0 \pm 0.1$                  \\ \cline{3-7} 
                                &                                     & Gemma 2                                 & $99\pm2$                                     & $2020 \pm 20$                                              & $98 \pm 1$                                                  & $7.3 \pm 0.0$                  \\ \cline{3-7} 
                                &                                     & Qwen 2.5                                & $81\pm4$                                     & $1988 \pm 61$                                              & $100 \pm 6$                                                 & $\mathbf{1.7 \pm 0.1}^\dagger$ \\ \cline{2-7} 
                                & \multirow{3}{*}{Qwen 2.5}           & Command-R                               & $99\pm2$                                     & $3854 \pm 240$                                             & $410 \pm 40$                                                & $15.3 \pm 1.0$                 \\ \cline{3-7} 
                                &                                     & Gemma 2                                 & $\mathbf{100\pm0}$                           & $4226 \pm 200$                                             & $398 \pm 29$                                                & $13.7 \pm 0.6$                 \\ \cline{3-7} 
                                &                                     & Llama 3.1                               & $99\pm2$                                     & $3977 \pm 169$                                             & $400 \pm 28$                                                & $5.5 \pm 0.3$                  \\ \hline

\multirow{13}{*}{T4}            & \multirow{3}{*}{Command-R}          & Gemma 2                                 & $\mathbf{73\pm0}$                            & $3003 \pm 187$                                             & $288 \pm 9$                                                 & $15.7 \pm 0.4$                 \\ \cline{3-7} 
                                &                                     & Llama 3.1                               & $68\pm4$                                     & $2808 \pm 184$                                             & $227 \pm 15$                                                & $8.5 \pm 0.2$                  \\ \cline{3-7} 
                                &                                     & Qwen 2.5                                & $72\pm2$                                     & $3030 \pm 51$                                              & $184 \pm 14$                                                & $8.3 \pm 0.1$                  \\ \cline{2-7} 
                                & \multirow{4}{*}{Gemma 2}            & Command-R                               & $70\pm4$                                     & $2692 \pm 191$                                             & $231 \pm 24$                                                & $14.6 \pm 1.0$                 \\ \cline{3-7} 
                                &                                     & Gemma 2                                 & $\mathbf{73\pm0}$                            & $2959 \pm 133$                                             & $284 \pm 9$                                                 & $10.1 \pm 0.3$                 \\ \cline{3-7} 
                                &                                     & Llama 3.1                               & $70\pm3$                                     & $2823 \pm 131$                                             & $981 \pm 1462$                                              & $6.6 \pm 5.3$                  \\ \cline{3-7} 
                                &                                     & Qwen 2.5                                & $\mathbf{73\pm0}$                            & $3062 \pm 148$                                             & $177 \pm 8$                                                 & $6.6 \pm 0.2$                  \\ \cline{2-7} 
                                & \multirow{3}{*}{Llama 3.1}          & Command-R                               & $55\pm2$                                     & $\mathbf{1652 \pm 41}$                                     & $\mathbf{93 \pm 10}$                                        & $6.9 \pm 0.3$                  \\ \cline{3-7} 
                                &                                     & Gemma 2                                 & $55\pm0$                                     & $1896 \pm 60$                                              & $106 \pm 5$                                                 & $6.7 \pm 0.1$                  \\ \cline{3-7} 
                                &                                     & Qwen 2.5                                & $55\pm0$                                     & $2168 \pm 188$                                             & $107 \pm 11$                                                & $\mathbf{1.9 \pm 0.2}^\dagger$ \\ \cline{2-7} 
                                & \multirow{3}{*}{Qwen 2.5}           & Command-R                               & $51\pm3$                                     & $2962 \pm 191$                                             & $241 \pm 27$                                                & $11.3 \pm 0.6$                 \\ \cline{3-7} 
                                &                                     & Gemma 2                                 & $64\pm0$                                     & $3348 \pm 280$                                             & $257 \pm 25$                                                & $10.9 \pm 0.6$                 \\ \cline{3-7} 
                                &                                     & Llama 3.1                               & $61\pm4$                                     & $2870 \pm 361^\dagger$                                     & $225 \pm 22$                                                & $3.3 \pm 0.3$                  \\ \hline

\multirow{13}{*}{T5}            & \multirow{3}{*}{Command-R}          & Gemma 2                                 & $\mathbf{55\pm2}$                            & $4799 \pm 226$                                             & $260 \pm 17$                                                & $15.6 \pm 0.4$                 \\ \cline{3-7}
                                &                                     & Llama 3.1                               & $45\pm3$                                     & $4622 \pm 124$                                             & $269 \pm 21$                                                & $9.5 \pm 0.2$                  \\ \cline{3-7} 
                                &                                     & Qwen 2.5                                & $44\pm3$                                     & $4985 \pm 222$                                             & $311 \pm 37$                                                & $10.2 \pm 0.5$                 \\ \cline{2-7} 
                                & \multirow{4}{*}{Gemma 2}            & Command-R                               & $37\pm4$                                     & $3735 \pm 77$                                              & $198 \pm 8$                                                 & $13.6 \pm 0.3$                 \\ \cline{3-7} 
                                &                                     & Gemma 2                                 & $38\pm2$                                     & $3546 \pm 87$                                              & $217 \pm 7$                                                 & $8.3 \pm 0.3$                  \\ \cline{3-7} 
                                &                                     & Llama 3.1                               & $28\pm2$                                     & $\mathbf{3325 \pm 101}$                                    & $\mathbf{197 \pm 9}$                                        & $\mathbf{3.5 \pm 0.1}$         \\ \cline{3-7} 
                                &                                     & Qwen 2.5                                & $29\pm3$                                     & $3942 \pm 60$                                              & $224 \pm 8$                                                 & $7.2 \pm 0.1$                  \\ \cline{2-7} 
                                & \multirow{3}{*}{Llama 3.1}          & Command-R                               & $12\pm5$                                     & $3614 \pm 276$                                             & $848 \pm 1240$                                              & $13.7 \pm 4.7$                 \\ \cline{3-7} 
                                &                                     & Gemma 2                                 & $20\pm8$                                     & $3598 \pm 208$                                             & $841 \pm 1234$                                              & $12.4 \pm 4.5$                 \\ \cline{3-7} 
                                &                                     & Qwen 2.5                                & $18\pm6$                                     & $3703 \pm 254$                                             & $877 \pm 1235$                                              & $5.9 \pm 4.4$                  \\ \cline{2-7} 
                                & \multirow{3}{*}{Qwen 2.5}           & Command-R                               & $28\pm3$                                     & $3487 \pm 342$                                             & $434 \pm 54$                                                & $13.3 \pm 1.3$                 \\ \cline{3-7} 
                                &                                     & Gemma 2                                 & $25\pm2$                                     & $5225 \pm 469$                                             & $520 \pm 60$                                                & $15.1 \pm 1.3$                 \\ \cline{3-7} 
                                &                                     & Llama 3.1                               & $26\pm4$                                     & $4128 \pm 438$                                             & $1065 \pm 1222$                                             & $7.9 \pm 4.4$                  \\ \hline
\end{longtable}

\end{document}